\documentclass[a4paper,reprint,superscriptaddress,secnumarabic,amsmath,amssymb,aps]{revtex4-1}

\usepackage{graphicx,xcolor}
\usepackage{ulem}
\graphicspath{{Figures/}}

\begin{document}

\title[Pressure and iodine defects in MAPI]{Effect of pressure on the dynamics of iodine defects in MAPI: An atomistic simulation}

\author{Rachel Elizabeth Brophy}
\affiliation{Department of Engineering, Reykjavik University, Menntavegur 1, IS-102 Reykjavik, Iceland}

\author{Movaffaq Kateb}
\affiliation{Department of Engineering, Reykjavik University, Menntavegur 1, IS-102 Reykjavik, Iceland}
\affiliation{Condensed Matter \& Materials Theory Division, Department of Physics, Chalmers University of Technology, SE-412 96 Gothenburg, Sweden}

\author{Kristinn Torfason} 
\affiliation{Department of Engineering, Reykjavik University, Menntavegur 1, IS-102 Reykjavik, Iceland}

\author{George Alexandru Nemnes} 
\affiliation{Research Institute of the University of Bucharest (ICUB), Mihail Kogalniceanu Blvd 36-46, 050107 Bucharest, Romania}
\affiliation{University of Bucharest, Faculty of Physics, 077125 Magurele-Ilfov, Romania}
\affiliation{Horia Hulubei National Institute for Physics \& Engineering, 077126 Magurele-Ilfov, Romania}

\author{Halldor Gudfinnur Svavarsson} 
\affiliation{Department of Engineering, Reykjavik University, Menntavegur 1, IS-102 Reykjavik, Iceland}

\author{Ioana Pintilie}
\affiliation{National Institute of Materials Physics, Atomistilor 405A, 077125 Magurele, Romania}

\author{Andrei Manolescu}
\affiliation{Department of Engineering, Reykjavik University, Menntavegur 1, IS-102 Reykjavik, Iceland}

\begin{abstract}
The diffusion of iodine defects has been considered the most important degradation mechanism of methylammonium lead iodine (MAPI) in solar cells. The present study demonstrates the importance of the pressure inside this material on the dynamics of iodine defects, using molecular dynamics simulations. It is known that the diffusion coefficient of an iodine vacancy is an order of magnitude higher than that of interstitial iodine. We show that this difference systematically increases with increased tensile strain and that both diffusion coefficients tend to zero when a compressive strain is applied.   This result suggests that compression of the MAPI can be a good solution to reduce its degradation rate. 
Besides, the statistical aspect of deriving the diffusion coefficient from the mean squared displacement (MSD) is discussed in terms of the initial conditions (positions and velocities) of the atoms and the simulation time, considering different seeds of the pseudo-random number generator used in the simulations performed with the LAMMPS software. 
\end{abstract}

\maketitle

\section{Introduction}
Methylammonium lead iodide (MAPI) \cite{kojima2009} is a promising photovoltaic material mainly due to its low-cost fabrication and high power conversion efficiency,  exceeding 24\% \cite{nrel}. It has a perovskite structure with CH$_3$NH$_3$PbI$_3$ stoichiometry in which the methylammonium (MA) molecule, or CH$_3$NH$_3$, has 1:1 ratio with Pb, and is located in the middle of a nearly L1$_2$ cell made of PbI$_3$. Then the structure polymorphism is dominated by the kinetics of MA molecules i.e.\ their random rotation gives cubic phase (Pm$\bar{3}$m) while more aligned molecules lead to tetragonal (I4/mcm) and orthorhombic (P4/mbm) phases. However, the MAPI material is currently characterised by the lack of stability 
and relatively short lifetime in the working conditions of a solar cell. It has been found that diffusion of iodine defects is the primary degradation mechanism of MAPI cell \cite{Sanchez14,eames2015,Li16,Besleaga16}. 
Whereas during the regular working regime, the iodine migration and accumulation at the interfaces with the charge transporter layers leads to a hysteretic behavior of the 
current-voltage characteristic of the device \cite{Tress15,Zarazua16,Nemnes17a}.

In order to describe the behavior of the MAPI cells with atomistic simulations, and in particular the iodine migration, a simple, but useful interatomic force field, called MYP, has recently been developed \cite{mattoni2015}. This force field allows modeling several thermodynamic properties of MAPI via molecular dynamics simulations (MDS) \cite{mattoni2017}. Using the MYP potential, the variation of defect dynamics with temperature could be simulated \cite{delugas2016}, the motion of defects near grain boundaries was compared with experimental data \cite{phung2020}, and strain introduced by an external electric field due to caloric effect has been studied \cite{liu2016}.
It has been shown that hydrostatic pressure can affect optical, electronic, and photovoltaic properties of various hybrid perovskites \cite{kong2016,jaffe2016,jaffe2017,liu2017,wang2017,postorino2017,ghosh2019}. 
Earlier, it had been shown that the diffusion barrier changes with variation of the lattice constant, which can be achieved by applying an external strain, or by alloying MA and I sites \cite{yang2016}.

In the present study, we utilize the MDS method to determine the effect of pressure on the iodine defect dynamics in MAPI. We consider separately the motion of an iodine vacancy, and of an interstitial iodine ion.  Apart of the presence of the defect, the MAPI material is assumed to be homogeneous.

\section{Methods}
We performed our simulation with the LAMMPS software package \cite{plimpton1995} which solves Newton's equation of motion based on parameterized interatomic forces, which were derived from the MYP potential \cite{mattoni2015}. The intra-MA interactions are modeled with a bounded generalized Amber force field (GAFF). Pb-Pb, I-I, Pb-I, and Pb/I-C/N interactions are considered non-bonded and modeled with Buckingham force field. Pb/I-H interactions were modeled by Lennard-Jones potential. Partial charges are assigned to each atom species and associated with pairwise Coulomb potentials. The Coulomb interaction was limited by a 10~{\AA} cutoff, and for the residual long-range contribution,  we utilized the Ewald summation with particle-particle particle-mesh (pppm) solver and accuracy of $10^{-4}$.

\begin{figure}
    \centering
    \includegraphics[width=0.40\linewidth]{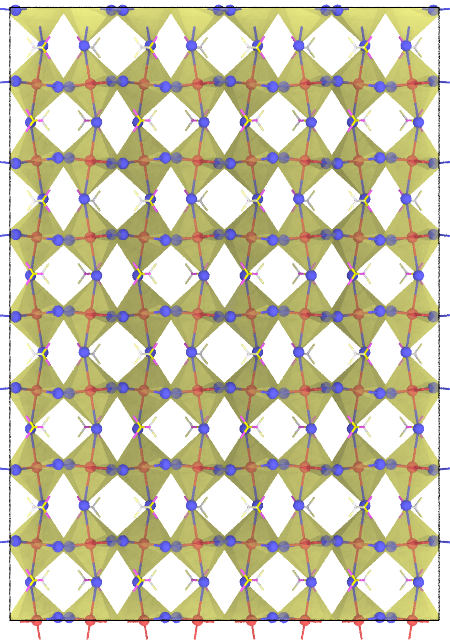}
    \includegraphics[width=0.57\linewidth]{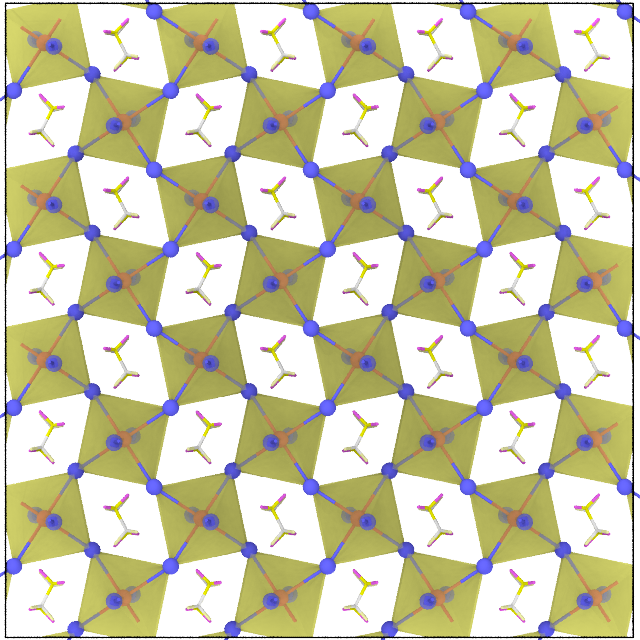}
    \caption{(left) $b$- and (right) $c$-axis views of the initial structure of defect-free MAPI. Each Pb atom in red is surrounded by 6 I atoms that together form octahedra indicated by yellow. Note, how MA molecules are oriented within empty spaces between octahedra in the c-axes view.}
    \label{fig:init_perfect}
\end{figure}

The initial defect-free MAPI structure was considered to be orthorhombic, with a simulation cell consisting of 3072 atoms which correspond to 
$8\times8\times4=256$ cubic unit cells. We assumed the $c$-axis of the orthorhombic crystal in the $z$ direction which leaves the basal plane ($ab$) of the crystal parallel to the $xy$ plane,  as shown in Figure~\ref{fig:init_perfect}. 
%
%


The system was relaxed in three steps. First, an energy minimization was performed using the conjugate gradient algorithm with a tolerance of $10^{-6}$~Kcal/mole{\AA} for the force, and $10^{-8}$ relative tolerance for the energy ($\Delta E/E$). 
We allowed the box dimension to vary during the minimization to produce the desired pressure, between -5$\times10^{3}$ and $10^{5}$~bar. Further relaxation was performed using the isothermal-isobaric (NPT) ensemble, by performing a temperature and pressure rescaling at each 100-th and 500-th time step (with every single step of 0.5~fs), using the Noose-Hoover thermostat and barostat, respectively. We gradually increased the temperature from 1 to 300~K within a time interval of 0.25~ns, and maintained it constant for an interval of 1.75~ns. 
Then, the NPT ensemble was used for the subsequent part of the simulation.  
The typical computational time was about 22 CPU hours for each nanosecond on a high-performance computer cluster of AMD EPYC processors 2300 MHz.

In practice, MAPI suffers from a relatively high density of various defects originating in the preparation process, most often in the original chemical solution used.  
In the present work, we consider two types of iodine defects: iodine vacancy and iodine interstitial. They can be generated simply by removing or adding an iodine to the lattice, respectively. They are both considered positively and negatively charged, respectively, and therefore, theoretically, these defect models may slightly violate the charge neutrality of the entire system. However, if the total number of atoms is sufficiently large, there should be no practical consequence on the atomistic simulations \cite{delugas2016}.  Instead, a paired iodine-vacancy defect can also be created by displacing an iodine from its original site to a desired distance. This model allows for maintaining the charge neutrality of the system, but it may not be viable for a long time as the pair may recombine.


In order to observe the diffusion of the iodine defects, we compute the time-dependent mean square displacement (MSD) of all iodine atoms during the simulation, which is
\begin{equation}
{\rm MSD}(t)=\frac{1}{N_I}\sum_{i=1}^{N_I} |{\bf r}_i(t) - {\bf r}_i(0)|^2 \ ,
\label{msddef}
\end{equation}
where $N_I$ is the total number of all iodine atoms in our simulation cell, and ${\bf r}_i(t)$ is the position of each iodine atom, $i=1,2,...,N_I$, at time $t$.  In the absence of mobile atoms, the MSD varies only due to the local thermal vibration, with a small average over the entire simulation time.   
Instead, the presence of mobile defects leads to a linear trend of MSD versus time, with a positive slope.  
Using the random walk theory the diffusion coefficient $D$ of the mobile defect can be calculated as \cite{allen1989}
\begin{equation}
D = \frac{1}{6}\lim_{t\to \infty} \frac{ {\rm MSD}(t) } {t} \, .
\label{diffcoeff}
\end{equation}
In practice $D$ can be obtained as the slope coefficient of the MSD versus time, divided by the factor 6, which is the double of the spatial dimension \cite{delugas2016,mattoni2017}.  The relaxation time after the initial positions, plus an additional extra time after that, is excluded from the slope calculation. 

We utilized the open visualization tool (OVITO) and its Python interface for post-processing and visualization.

\section{Results}
\subsection{Iodine vacancy}
\begin{figure}
    \centering
    \includegraphics[width=1.0\linewidth]{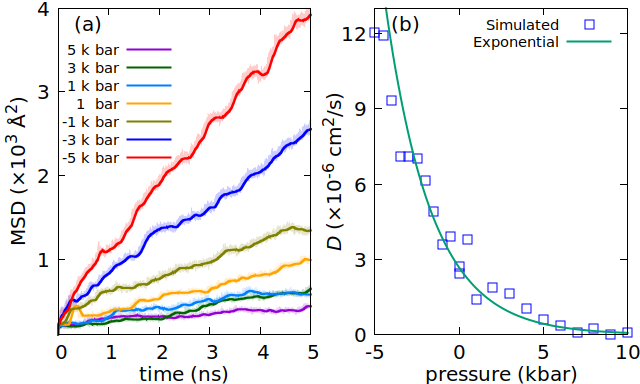}
    \caption{(a) MSD of iodine vacancy at 300~K and different pressures. The lines show a smeared MSD using 2000 time points, while the original data is shown as a shaded area around the lines. (b) Variation of $D$ with pressure, for more values of the pressure, the solid line being the exponential fit mentioned in the main text.}
    \label{fig:msd_Ivac}
\end{figure}

Fig.~\ref{fig:msd_Ivac} shows the time-dependent MSD and the corresponding diffusion coefficients $D$ for an iodine vacancy at 300~K and different applied pressures. 
It can be seen that the negative pressures, corresponding to a tensile strain (and an expansion of the material), lead to an increased slope of the MSD, and of the corresponding diffusion coefficient, while in the presence of a  positive pressure (or a compression of the material), the diffusion coefficient decreases, and eventually tends to zero at about 10~kbar (not shown).

The graph clearly indicates an exponential decrease of $D$ with increasing pressure $p$. A linear fit of the resulting values of the diffusion coefficient of the iodine vacancy, on the log scale, yielded an R-squared coefficient of 0.92.  The corresponding solid line shown in Fig.\ \ref{fig:msd_Ivac}(b) is the resulting fitting function, $2.67\times 10^{-6}\exp(-0.364 p)$.
At atmospheric pressure our diffusion coefficient is $2.42\times 10^{-6}$ cm$^2$/s, close to the value reported by Delugas et al., which is 4.3$\times10^{-6}$~cm$^2$/s, obtained at 300~K, but using the canonical thermodynamic ensemble, NVT, implemented in the DL\_POLY molecular simulation package \cite{delugas2016}. The decrease of the MSD and of the diffusion coefficient with increasing pressure can be associated with the increase of the density of the MAPI material, leading to a blocking of the iodine atoms at their original sites, or to an increase of the energy they need for jumping into the nearby vacancy.    

As we can see in Fig. \ref{fig:msd_Ivac}(a), the instantaneous time derivative of the MSD, $d({\rm MSD})/dt$, is not a smooth function of time. For this reason, for a reasonably accurate estimation of the diffusion coefficient $D$, a simulation time longer than the average oscillation time of the MSD slope is needed, and apparently, a total time of several ns should be sufficient.  However, the observable variation of the slope, over shorter time intervals, also suggests that the MSD evolution may depend on the initial conditions.  To check that we repeated the simulations by changing the seed of the (pseudo)random number generator in LAMMPS, which implies different initial states of the entire lattice at the beginning of the simulation, and also after the relaxation phase, and consequently different trajectories of the entire system in the phase space. In addition, we also used different (random) initial placements for the vacancy.  

\begin{figure}
    \centering
    \includegraphics[width=1.0\linewidth]{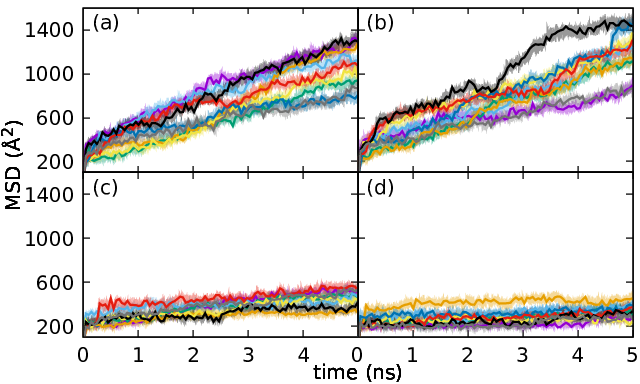}
    \caption{(a) The MSD in the presence of an iodine vacancy, at pressure 1 bar, using different seeds of the random number generator, and (b) using also different initial positions of the vacancy. (c,d) The same at pressure 5 kbar.  The temperature was fixed at 300 K. 
    }
    \label{fig:seed_Ivac_msd}
\end{figure}

Fig.~\ref{fig:seed_Ivac_msd}(a,c) shows the MSD results for the iodine vacancy with different seeds at 1~bar and 5~kbar, respectively. In both cases, the initial positions of all atoms were the equilibrium positions plus random displacements with $-0.1<\delta<0.1$~{\AA} in each direction $x,y,z$, depending on the seed. The initial velocities of the atoms were also assigned randomly, from a Gaussian distribution, again depending on the seed. Thus, all simulations began with different initial conditions. Additionally, in Fig.~\ref{fig:seed_Ivac_msd}(a,c) the iodine vacancy had the same initial position, whereas in Fig.~\ref{fig:seed_Ivac_msd}(b,d) this position was also randomized. 
Clearly, the effect of the seed creates significant variations of the slope. This is evident at both pressures used. We can also see that at the higher pressure, of 5~kbar, the slope is systematically reduced, in agreement with the previous result that the diffusion of iodine vacancies is suppressed at elevated pressures. 

In the nine simulations represented in Fig.~\ref{fig:seed_Ivac_msd}~(a), at 1~bar, the slope coefficients are, in increasing order: 96, 104, 151, 159, 163, 179, 189, 204, 229~\AA$^2$/ns, with a mean value of 164~\AA$^2$/ns and a standard error (or statistical uncertainty) of about 14~\AA$^2$/ns, meaning $\pm$9\% relative uncertainty for the estimated diffusion coefficient.  With the data obtained at 5~kbar the relative uncertainty of $D$ is about $\pm$15\%. 
At the same time, the uncertainty of the slope coefficient corresponding to a single simulation is much smaller, in the range 0.1--0.3\%, or less, because it is based on a sample with millions of time points. 

Therefore, it turns out that every single run of several ns does not include sufficient information about the distribution of all possible states of the system in the phase space. This is true because the slope coefficients obtained with 4~ns simulations differ from their mean values obtained with different initial conditions.  
It is also seen in our figures that the slope of the MSD vs.\ time changes on intervals of the order of 1~ns or less. Similar slope variations can also be observed in the calculations of Delugas et al., in their Figure~3, representing the MSD data for iodine vacancy and iodine interstitial at different temperatures  \cite{delugas2016}.  
Our interpretation is that a computational time of several ns is not sufficient to observe the ergodic behavior of our system, i.e.\ the convergence of the time-averaged and phase space-averaged values of the diffusion coefficient.  For that purpose an unrealistically long computational time might be necessary, or, instead, a reasonable accuracy could be achieved with shorter simulations with different initial conditions. Long-time memory of initial conditions in a diffusion process has also been addressed by other studies and formulated on a rigorous mathematical background \cite{Banerjee22}. However, such fundamental aspects of statistical mechanics are beyond the scope of our present work, rather based on an empirical approach.

\subsection{Iodine interstitial}

\begin{figure}
	\centering
 \includegraphics[width=1.0\linewidth]{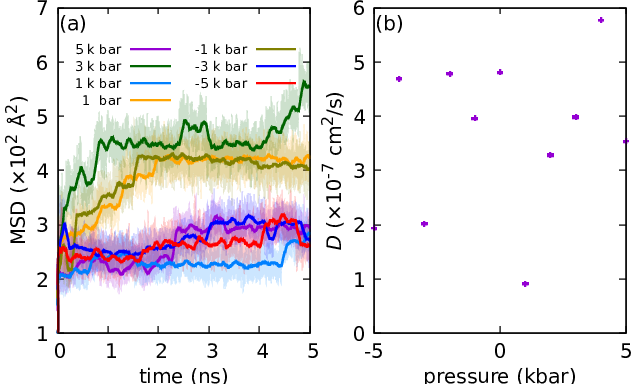}
	\caption{(a) MSD of charged iodine interstitial at 300~K and different pressures. Darker lines indicate a moving average of 2000 points while the original data are shown with lighter colors in the background. (b) Variation of $D$ with pressure. }
	\label{fig:msd_Iintch}
\end{figure}

Now we consider an interstitial iodine ion (i.e.\ an iodide) added to the MAPI lattice, and repeat the previous simulations. 
The results are shown in Fig.~\ref{fig:msd_Iintch}, obtained with the NPT ensemble, at different pressures.  As before, the MSD corresponds to all iodine atoms, and the diffusion coefficient of the interstitial ion is found from the slope of the MSD. Note that the scale of the diffusion coefficient $D$ is an order of magnitude smaller than in the case of the vacancy,  Fig.~\ref{fig:msd_Ivac}(b). Again, we are close to the value previously reported by Delugas et al., of 7.4$\times10^{-7}$~cm$^2$/s, obtained using the NVT ensemble at 300~K \cite{delugas2016}. 
We also find out that the diffusion coefficient of the interstitial iodide is at least one order of magnitude smaller than that of the vacancy at atmospheric pressure. 

However, in the iodide case, we could not find a systematic variation of the diffusion coefficient with the applied pressure.  
The values shown in Fig.~\ref{fig:msd_Iintch}(b) are scattered, and probably not convergent after 5~ns simulation time. The slope of the MSD data now looks variable over larger time intervals than in the case of the vacancy. And, as expected, by repeating the simulations with different seeds of the random number generator, the overall slope of MSD with respect to time is systematically smaller than in the case of the iodine vacancy, and possibly with a larger uncertainty.  

\begin{figure}
	\centering
\includegraphics[width=1.0\linewidth]{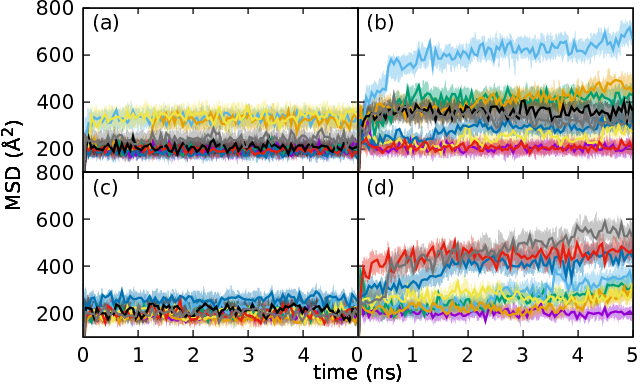}
	\caption{Variation of the MSD for the negatively charged iodine (iodide) interstitial with the seed of the random numbers. (a,b) correspond to 1 bar and (c,d) to 5 kbar pressure. In (a) and (c) the initial position of the iodide was the same, whereas in (b) and (d) it was randomized.}
	\label{fig:seed_Iintch_msd}
\end{figure}

The results obtained with different seeds, for pressures of 1 bar and 5 kbar,  are shown in Fig.~\ref{fig:seed_Iintch_msd}.
It is clearly seen that when a defect is introduced in the same place, Fig.~\ref{fig:seed_Iintch_msd}~(a,c), the MSD data look similar and nearly flat. 
However, as can be seen in Fig.~\ref{fig:seed_Iintch_msd}~(b,d), introducing the defect elsewhere might lead to different MSD data, but their slope being always much smaller than in the case of the vacancy.  Nonetheless, the pressure seems to have a minor effect in these MSDs, indicating that iodide diffusion is a less frequent event, at least up to a pressure of 5~kbar, which was our largest value.

\begin{figure}
    \centering
    \includegraphics[width=1.0\linewidth]{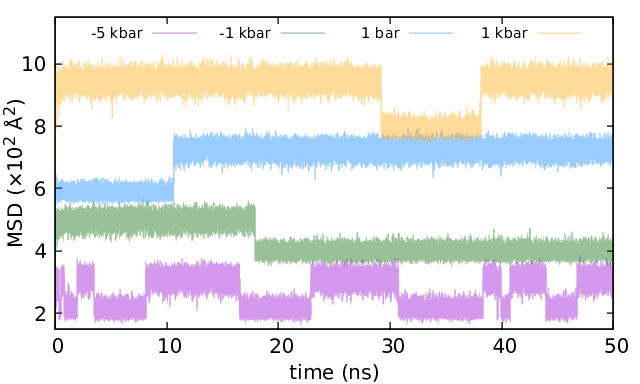}
    \caption{MSD of iodine interstitial at different pressures obtained over 50~ns time scale. The curves are shifted manually for clarity.}
    \label{fig:Iintch_msd_50ns}
\end{figure}

To clarify these results, ten times longer simulations were performed for the case with the interstitial iodide, at four different pressure values.  The resulting MSDs are shown in Fig.~\ref{fig:Iintch_msd_50ns}. Note that the results corresponding to different pressures are shifted vertically, for illustration purposes. At this time scale, for all pressure values, we can only identify jumps of the MSD, on the background of all iodine atoms' back-and-forth movements or local vibrations. The jumps are of approximately $\sqrt{\rm MSD}\approx \sqrt{100} = 10$~\AA, which correspond to the size of a unit cell. So at this time scale, no net motion of the iodide can be identified, but only hopping events from one unit cell to another, which appear to be more frequent at high negative pressures such as -5~kbar, than at positive pressures.


\section{Conclusions}

We performed molecular dynamics simulations of the iodine migration in the MAPI material, using the LAMMPS software, over time intervals from a few to a few tens of ns, and we obtained estimated values of the diffusion coefficients of the iodine vacancies and interstitial atoms, for different negative and positive pressures applied to the MAPI material.  
Our simulations show that the iodine vacancy has a diffusion coefficient, or, equivalently, a mobility, that depends significantly on the pressure. The mobility of the vacancy varies exponentially with the pressure, increasing in the presence of tensile stress (negative pressure compared to the atmospheric value), and decreasing in the presence of compressive stress (positive pressure). For the same conditions, the mobility of an interstitial iodine negative-ion (or iodide) is at least one order of magnitude lower, which means that the net, macroscopic, iodine migration can be associated with the diffusion of the iodine vacancies. 
Therefore the migration of the iodine atoms, which is a major factor in the degradation of the MAPI material, could be suppressed, or at least reduced, by applying to the material a compressive stress of the order of a few kbar.            

Recent studies also suggested that with even larger pressures, of the order of hundreds of kbar, the band gap can also be modified, although achieving such pressures could be technically difficult, and could possibly lead to phase transitions of the MAPI material \cite{Rajeswa20,Faghihna20,Lee20}. 

In our simulations, we assumed a homogeneous MAPI lattice and no external electric field. The presence of lattice defects such as grain boundaries can change the physical conditions.  The local electric field may attract the ionic defects towards the grain boundaries \cite{phung2020}, inhibiting further their diffusion.  However, the iodine migration may still occur at the contact between the MAPI material and another layer, for example, the hole transporter made of Spiro-OMeTAD \cite{Besleaga16}.

\section{Appendix. Time-dependent fluctuations of the diffusion coefficient}

In a more rigorous treatment than ours, the single atom diffusion coefficient is given by the velocity auto-correlation function \cite{allen1989}
\begin{equation}
	D_i=\frac{1}{3}\int_0^\infty{\rm d}t\Big\langle v_i(t).v_i(0)\Big\rangle \ ,
\end{equation}
where 3 is the number of degrees of freedom (the 3D physical space), $v_i(t)$ velocity of particle $i$, and the angular brackets denote the statistical average.
However, in numerical implementations, this method gives some level of background noise fluctuations. Therefore, the corresponding Einstein relation that can be valid at a very long time is preferred \cite{allen1989}, which leads to
\begin{equation}
	2tD_i=\frac{1}{3}\Big\langle|\mathbf{r}_i(t)-\mathbf{r}_i(0)|^2\Big\rangle_t \ ,
\end{equation}
where $\mathbf{r}_i(t)$ is the particle position. Here, the angular brackets correspond to the time-averaged mean-squared displacement, which is another form of Eq.\ (\ref{diffcoeff}) from the main text.

In practice, there exist two approaches for the calculation of the diffusion coefficient for defects. The first approach requires the calculation of the coordination number. Then a vacancy or interstitial is associated with a dummy particle located as the center of mass of the under- and over-coordinated atoms, respectively. For instance, Pb has a coordination number of six within the defect-free MAPI structure. Thus, two under- and over-coordinated Pb (i.e.\ with 5 or 7 neighbors, respectively) represent an iodine vacancy and interstitial, respectively \cite{delugas2016}. 
This method, however, is very sensitive to the definition of cutoff for finding nearest neighbors and lattice vibration may introduce a considerable noise in determining coordination number. Thus the second approach is based on the existing particles rather than a dummy one. For example, the movement of an iodine vacancy is associated with the hopping of the nearby iodines into it. Thus, it is possible to calculate the MSD of the defect through the MSD of all iodine atoms. It is worth mentioning that the MSD can be summed over a group of particles and averaged over the number of particles. For the defect these averaged values must be multiplied by the number of particles \cite{nist} i.e. 
\begin{equation}
	{\rm MSD}(t)_{\rm defect}=N\times {\rm MSD}(t)_{\rm total} \ ,
\end{equation}
with $N$ being total number of atoms defining the defect, which in our study are the iodine atoms.

\vspace{12mm}

\begin{figure}[h]
	\centering
\includegraphics[width=1.0\linewidth]{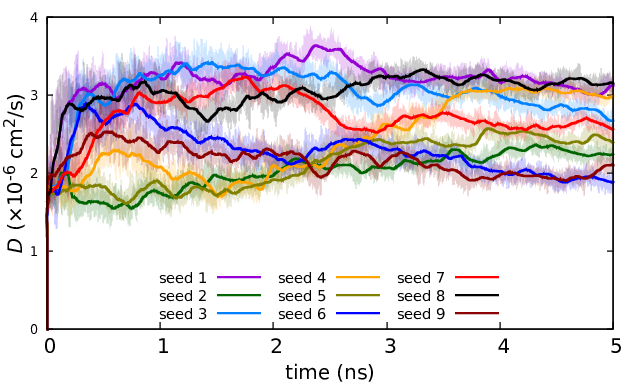}
	\caption{Instantaneous diffusion coefficient for different seeds and a fixed position of the iodine vacancy at 300~K and 1~bar. The darker lines show a moving average of 2000 points while the data is plotted with lighter colors.}
	\label{fig:seed_Ivac_d2p}
\end{figure}

\clearpage

In Fig.\ \ref{fig:seed_Ivac_d2p} we show the time-averaged values of the $D$ coefficient for the iodine vacancy, averaged over 2000 time steps, where we can see the fluctuations during each simulation, and also for different initial conditions imposed by the initial values of the random number generator.  These graphs show again that at this time scale the diffusion coefficient of vacancies has a distribution whose standard deviation is about 30\%-50\% of the mean value.  It is (presumably) a normal distribution because each $D$ value has been obtained as a sample mean.  


\begin{acknowledgments}
The research leading to these results has received funding from the EEA Grants 2014–2021, under Project contract no. 36/2021 (project code: EEA-RO-NO-2018-0106) and from the Core Program 2019–2022 (contract 21N/2019).  REB is thankful for partial support from the National Power Company of Iceland (Landsvirkjun) via the Sustainability Institute and Forum (SIF) of Reykjavik University.  The computational resource was sponsored by EGI and the EGI-ACE H2020 project (GA no. 101017567) with the dedicated support of CLOUDIFIN.
\end{acknowledgments}

\bibliography{Ref}

\end{document}